\documentclass[prl,twocolumn,showpacs]{revtex4}

\usepackage{latexsym,amsmath,graphics,graphicx}

\newcommand{\vc}{\mathbf}
\newcommand{\ct}{\cite}
\newcommand{\hhf}{H_{\mathrm{hf}}}

\begin{document}

\title{Coordination Dependence of Hyperfine Fields of $5sp$
Impurities on Ni Surfaces} 
\author{Phivos Mavropoulos} 
\affiliation{Institut f\"ur Festk\"orperforschung, Forschungszentrum
J\"ulich, D-52425 J\"ulich, Germany}
\date{\today}

%%%%%%%%%%%%%%%%%%%%%%%%%%%%%%%%%%%%%%%%%%%%%%%%%%%%%%%%%%%%%%%%%%%%%%
\begin{abstract}
  We present first-principles calculations of the magnetic hyperfine
  fields $\hhf$ of $5sp$ impurities on the (001), (111), and (110)
  surfaces of Ni. We examine the dependence of $\hhf$ on the
  coordination number by placing the impurity in the surfaces, on top
  of them at the adatom positions, and in the bulk.  We find a strong
  coordination dependence of $\hhf$, different and characteristic for
  each impurity. The behavior is explained in terms of the on-site
  $s$-$p$ hybridization as the symmetry is reduced at the surface. Our
  results are in agreement with recent experimental findings.
\end{abstract}
%%%%%%%%%%%%%%%%%%%%%%%%%%%%%%%%%%%%%%%%%%%%%%%%%%%%%%%%%%%%%%%%%%%%%%


\maketitle

The understanding of the magnetic properties of solids on the atomic
level is a challenge for experiment and theory. The magnetic hyperfine
interaction of the electronic magnetization with the nuclear magnetic
moment of an atom is such a property, and it is known to depend in
turn on the properties of the atom as well as on the neighboring
atomic environment. In particular, for impurities in the bulk of the
ferromagnetic materials Fe, Co, Ni, the trends of magnetic hyperfine
field $\hhf$ have been well understood and the experimental and
theoretical results extend practically over the whole periodic table
\ct{Rao85,Akai90}. The relation of $\hhf$ to the local magnetic moment
and to the atomic environment is not trivial. Therefore, for the
interpretation of the trends, first-principles, all-electron methods
have been extremely useful \ct{Akai90}, since they can describe the
charge density and magnetization near the atomic nucleus
self-consistently.  $\hhf$ measurements for probe atoms at interfaces
or surfaces can also provide unique information on the local structure
\ct{Bertschat99}, especially in combination with relevant calculations
\ct{Bellini01,Rodriguez01,Mavropoulos98a}.

One of the modern powerful experimental techniques allowing the study
of the hyperfine interaction of impurities at surfaces or interfaces
is the perturbed angular correlation spectroscopy \ct{Pleiter82}. It
has a high enough sensitivity that allows accurate measurements from
highly diluted probe atoms ($10^{-4}-10^{-5}$ of a monolayer), so that
these form practically isolated impurities. In a recent experimental
publication \ct{Potzger02}, the coordination number dependence of
$\hhf$ of Cd impurities in Ni host was studied, by comparing data from
Cd positioned on several Ni surfaces. The conclusion was that the
$\hhf$ of Cd strongly depends on the coordination number. Although
this might be expected, since the magnetic moment of the Cd impurity
is not intrinsic but rather induced by the environment, the data show
neither a linear dependence nor an increase with the number $N$ of Ni
neighbors. On the contrary, $\hhf$ is measured to be strongest for
$N=3$, while as $N$ is progressively increased $\hhf$ seems to change
sign and has a parabolic behavior.

In order to interpret this puzzling behavior, we have performed
first-principles calculations for all $5sp$ and the early $6s$
impurities (Ag to Ba) on various Ni surfaces and in bulk Ni in a
substitutional position. The impurity positions studied are given in
Table~\ref{tab:1} together with the coordination numbers. In this way
we aim to understand the trends of $\hhf$ not only on the coordination
number, but also on the impurity atomic number as has been done in the
past in the bulk and on (001) surfaces
\ct{Akai90,Mavropoulos98a,Mavropoulos98b}. In all cases the impurities
were assumed to be on the ideal lattice positions with the
experimental fcc Ni lattice constant; relaxations were not accounted
for.
\begin{table}[b]
\caption{Impurity positions on Ni surfaces and in bulk and their
coordination numbers $N$. $S$: impurity is in the surface layer;
$S+1$: impurity is on top of the layer in the hollow adatom position;
$S-1$: impurity is in the sub-surface layer; (111) Kink:
impurity is on top of (111) surface at a kink.}
\label{tab:1}
\begin{tabular}{lccccccccc}
\hline
Surf. & (111) & (001) & (110) & (111) & (110) & (001) & (111) & (110) & Bulk\\
Pos.  & $S+1$ & $S+1$ & $S+1$ & Kink  & $S$  &  $S$  &  $S$  & $S-1$ &     \\
\hline
$N$   &   3   &   4   &   5   &   6   &  7   &   8   &   9   &  11   &  12 \\
\hline
\end{tabular}
\end{table}

The calculations are based on the local spin density approximation of
density-functional theory. The full-potential Korringa-Kohn-Rostoker
Green function method for defects in bulk or at surfaces is employed
\ct{Papanikolaou02}, with an exact description of the atomic cells
\ct{Stefanou90}. In short, after calculating the electronic structure
of the host medium (bulk or surface) self-consistently, we use the
Green function of this reference system to calculate the electronic
structure of the distorted system containing the impurity via an
algebraic Dyson equation.  The power of the method lies in that it
works in real space with the boundary condition of the infinite host
included, rather than in $k$-space with a supercell construction. The
Green function $G^{nn'}_{lm;l'm'}(E)$ is described in cell-centered
coordinates around cells $n$ and $n'$ and expanded in cell-centered
solutions of the Schr\"odinger equation with angular momenta $(l,m)$
and $(l',m')$. The impurity and a cluster containing the 12 first
neighbors is perturbed in our calculations; this is enough because the
fcc crystal structure is close packed. Increasing the cluster to
contain the second neighbors in a test case brought insignificant
changes to our results. A truncation of the angular momentum to
$l_{\mathrm{max}}=3$ was taken. In a final step, the trace of the
Green function is used to obtain the $(l,m)$ decomposed charge density
and magnetic moment. We use the scalar relativistic approximation
\ct{Koelling77}, which takes into account the relativistic effects
other than the spin-orbit interaction and retains spin as a good
quantum number.

In the case of spin magnetism, the dominant contribution to $\hhf$ is
the Fermi contact interaction relating it directly to the spin density
at the nucleus, $m(\vc{r}=0)$. In a non-relativistic treatment this
has the simple form
\begin{equation}
\hhf = \frac{8\pi}{3} m(\vc{r}=0).
\label{eq:1}
\end{equation}
In a scalar relativistic treatment, as in the present paper, Breit's
formula has to be used instead to average $m(\vc{r})$ over the Thomson
radius of the nucleus \ct{Blugel87,Akai90}. In any case the
magnetization at or near the nuclear position is important, and
determined by the $s$ wavefunctions alone, since the states of higher
$l$ vanish at $\vc{r}=0$.

In $sp$ impurities, which possess no intrinsic magnetic moment, the
magnetization is transferred by the neighboring magnetic atoms. This
has almost no effect on the contribution to $\hhf$ from $s$ bound
states of the ionic core, but affects the contribution from the
valence states strongly. This is found in our calculations, in
agreement to previous results \ct{Akai90}; in fact, the valence
contribution to $\hhf$ is proportional to the local $s$ moment
\ct{Akai90,Mavropoulos98b}. Thus we focus on the behavior of the
valence $s$ states.

\begin{figure}
\begin{center}
\includegraphics[angle=90,width=8cm]{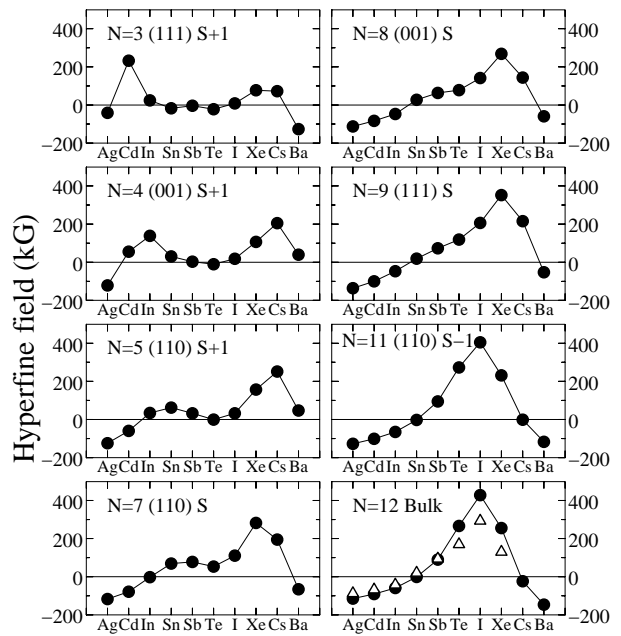}
\caption{Magnetic hyperfine field of $5sp$ impurities in Ni bulk and
  on various Ni surfaces as a function of the impurity atomic number
  $Z$. The coordination number $N$ depends on the surface and position
  (see Table~\protect{\ref{tab:1}}). As $N$ increases from 3 to
  12, the double-peaked structure of $\hhf(Z)$ transforms
  progressively to a single-peaked one. This is related to the gradual
  decoupling of the $s$ from the $p$ states as explained in the text.
  For $N=12$ the triangles show the experimental values
  \protect{\cite{Rao85}}.}
\label{fig:1}
\end{center}
\end{figure}

Our starting point is the interpretation of the trends of $\hhf$ as a
function of the impurity atomic number $Z$, as it has been understood
generally for defects in the bulk of ferromagnetic hosts in the past
\ct{Kanamori81,Akai90}. Relevant results are shown in Fig.~\ref{fig:1}
(bottom-right) together with experimental data taken from
Ref.~\cite{Rao85}. The central issue is the so-called $s$-$d$ {\it
  hybridization} of the impurity $s$ orbitals with the $d$ states of
the magnetic host, forming bonding (lower in energy) and
antibonding (higher) hybrid states. In this way the $s$ local density
of states (LDOS) of the impurity is affected, differently for every
spin due to the exchange splitting of the host $d$ band. We denote the
host majority spin as spin-up and the minority as spin-down. In the
beginning of the series (Ag) the bonding hybrids for both spins are
occupied while the antibonding ones unoccupied.  Because the host
spin-down $d$ band is higher in energy than the spin-up, the spin-down
bonding hybrids are more $s$-like than the spin-up ones.  Then the
impurity $s$ moment is negative and so is $\hhf$. As we change $Z$ to
the next impurities, the antibonding hybrids come to lower energies
and are progressively populated. First to cross the Fermi level $E_F$
is the spin-up antibonding hybrid, thus the $s$ moment and $\hhf$
increases and changes sign, peaking at Iodine. Then the spin-down
antibonding hybrid is gradually populated, the $s$ moment decreases
again, and so does $\hhf$.

Next we turn to cases of reduced coordination. Several coordinations
can be realized with the impurity at (001), (111), and (110) surfaces
as shown in Table~\ref{tab:1}. The results for $\hhf(Z)$ for these
coordination numbers $N$ are shown in Fig.~\ref{fig:1}. We see that,
as the coordination is gradually reduced, the single-peaked structure
of $\hhf(Z)$ evolves to a double-peaked one. Already for $N=8$ a
``shoulder'' appears at Sb, and for $N=5$, 4 and 3 there are clearly
two maxima of $\hhf$ with a local minimum of almost vanishing $\hhf$
in-between, at Te. A second observation is that, at the beginning of
the series, $\hhf$ changes sign to positive much earlier at the
surface that in the bulk --- for $N=4$ and 3 the change occurs already
between Ag and Cd, while in the bulk it occurs at Sn.

To interpret the results we note that the reduced coordination has
three effects. Firstly, the reduction in the transfered magnetic
moment, secondly the reduction in the strength of the $s$-$d$
hybridization, and thirdly the reduction of symmetry. As for the first
effect, a smaller impurity moment does not necessarily lead to a
reduced $s$ moment and weaker $\hhf$. Indeed, when the second effect
is also considered, a weaker $s$-$d$ hybridization leads to a smaller
bonding-antibonding splitting of the $s$-$d$ hybrids. The antibonding
hybrids (in particular for spin-up) appear lower in energy for smaller
$N$, and are populated earlier in the $sp$ series and also $\hhf$
increases earlier. This is consistent with the rise of $\hhf$ for Ag
and Cd impurities with reducing the coordination.

But there is still the important third effect, {\it i.e.}, the
reduction of symmetry, which results in an interaction of the $s$ with
the $p$ states that coexist around $E_F$ for $sp$ atoms. In the bulk
the impurity has a cubic environment, and the $s$ and $p$ orbitals
transform according to different irreducible representations, {\it
i.e.}, an eigenfunction cannot contain both kinds of orbitals
simultaneously. As the coordination number is reduced at the surface,
cubic symmetry no longer holds and the $s$ and $p$ orbitals are no
more decoupled. In the cases studied here, if we choose the $z$ axis
to be normal to the surface, it is actually the $p_z$ orbital that is
coupled with the $s$ (but in a more asymmetric case as at a step all
$p$ orbitals would couple). Then $s$-$p$ hybrids must exist, here in the
form $|s> \pm |p_z>$. The $s$ LDOS becomes more complicated, and in
particular each of the $s$-$d$ hybrids discussed above is expected to
split in two (one for each $s$-$p$ hybrid). Although the $s$-$p$
hybridization should happen even for $N=11$, it is stronger for
smaller coordination numbers when the environment becomes less
bulklike.

The new picture is as follows. With reduced coordination, two
antibonding $s$-$d$ hybrids exist per spin, one for each $s$-$p$
hybrid.  In the beginning of the $sp$ series (Ag) only bonding hybrids
are occupied and $\hhf$ is negative, similar to the bulk case. In the
next elements the first antibonding hybrid (which is more $s$-like
since the $p$ states are still high) crosses $E_F$, first for spin
up, leading to an increase of $\hhf$, and then for spin down, leading
to a decrease. Thus the first peak of $\hhf$ is formed (for $N=7$, 5,
4, and 3, at Sb, Sn, In, and Cd, respectively). As $N$ is reduced the
first peak moves toward the beginning of the series because the
$s$-$d$ hybridization is weaker and the first $s$-$d$ antibonding
states are lower in energy. Afterward the second antibonding hybrid
(now more $p$-like, since the $p$ states are lower) crosses $E_F$ and
becomes populated, again first for spin up (increasing $\hhf$) and
then for spin down (decreasing $\hhf$); the second peak in $\hhf$ is
thus formed. In this way the reduction in symmetry for smaller
coordination results in a double-peak structure in $\hhf$. The cases
for $N=8$ and 9 are intermediate, in which the $s$-$p$ hybridization
is weak and the first peak merges with the second one creating the
shoulder at Sb.

\begin{figure}
\begin{center}
\includegraphics[angle=90,width=8cm]{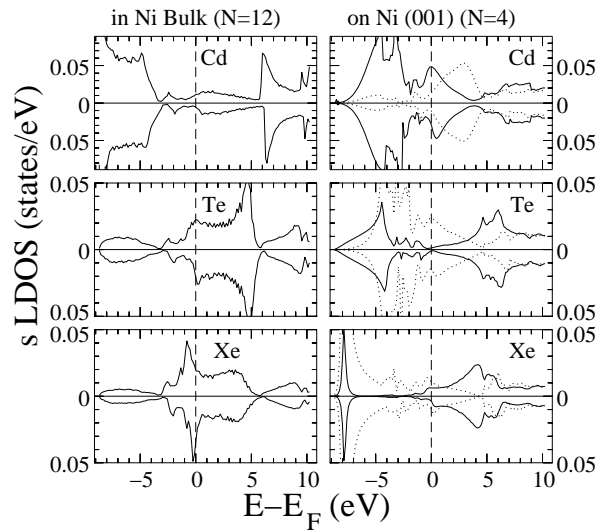}
\caption{Spin-resolved $s$ LDOS of Cd, Te, and Xe impurities in bulk
Ni (left) and on Ni (001) (right, full lines). For the adatoms, the
$p_z$ LDOS is also shown in dotted lines; for Te and Xe the $p_z$ LDOS
has been scaled by a factor of 0.5. Note also the different scale for
Cd compared to Te and Xe. Upper panels correspond to spin up,
lower panels to spin down.}
\label{fig:2}
\end{center}
\end{figure}
In Fig.~\ref{fig:2} the $s$ LDOS is shown for Cd, Xe, and Te
impurities in bulk Ni (N=12) and on top of the (001) Ni surface as
adatoms (N=4). In the adatom case also the $p_z$ LDOS is presented in
dotted lines. First we discuss the progressive filling of the states
for impurities in bulk. For Cd, the bonding $s$-$d$ hybrids are
occupied for both spins, and the spin-up antibonding hybrid is just
starting to pass through $E_F$; $\hhf$ starts to increase. For Te, the
antibonding hybrids are more occupied, and for Xe the spin-down
antibonding peak becomes occupied. Thus we have the single-peaked
structure of $\hhf(Z)$.  Now we turn to the adatom case. The $s$ LDOS
(full line) looks different. For Cd, the first antibonding hybrid is
at $E_F$, and is more pronounced than in the bulk due to the reduced
coordination and hybridization. The $p_z$ states (dotted line) are
energetically a little higher, but we can already recognize some
correlation between $s$ and $p$ at the peaks at $-1$ eV. $\hhf$ is
approaching its first maximum. Next, for Te, the first antibonding $s$
hybrid is populated for both spins, while the second is above $E_F$; a
``neck'' of almost zero $s$ LDOS is at $E_F$, and the $s$ moment and
$\hhf$ are almost zero. The $p$ states are lower, and an $s$-$p_z$
correlation is seen as the peaks coincide between $-5$ eV and $E_F$.
Finally, for Xe, the first antibonding hybrid is low (at $-8$ eV
accompanied by a peak in $p_z$), while the second is passing by $E_F$,
first for spin up giving the second rise to $\hhf$.

\begin{figure}
\begin{center}
\includegraphics[angle=90,width=8cm]{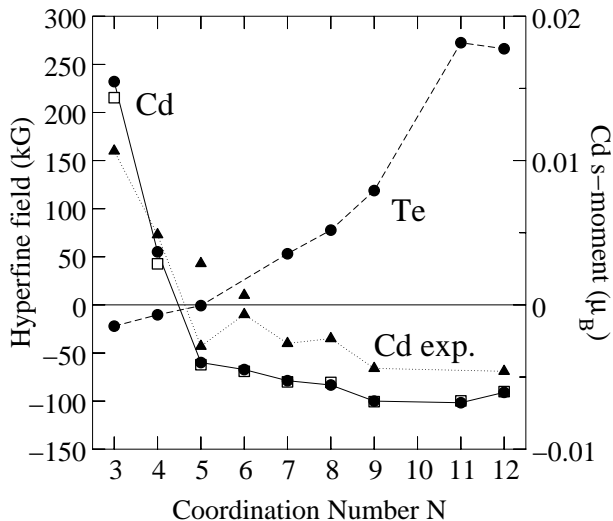}
\caption{Calculated coordination number dependence of $\hhf$ of Cd
(full line, full circles) and Te (dashed line) probe atoms on Ni
surfaces. The triangles correspond to experimental values reported in
Ref.~\ct{Potzger02}. The coordination number $N$ depends on
the surface and on the position as described in
Table~\protect{\ref{tab:1}}. While for Cd $\hhf$ increases for lower
$N$ as the spin-up antibonding $s$-$d$ hybrids come lower in energy, the
situation is opposite for Te, where the on-site $s$-$p$ hybridization
for low $N$ splits the $s$-$d$ hybrids in two, one populated and one
unpopulated, for each spin. The squares show the Cd impurity $s$
moment (right axis), to demonstrate its proportionality to the
$\hhf$.}
\label{fig:3}
\end{center}
\end{figure}
Now the observed coordination dependence of $\hhf$ of Cd probe atoms
can be understood. In Fig.~\ref{fig:3}, $\hhf(N)$ is shown, together
with the experimental results \ct{Potzger02} and the calculated
results for Te probes. For Cd $\hhf$ increases and changes sign for
lower $N$ as the spin-up antibonding $s$-$d$ hybrids come lower in
energy, because of the weaker $s$-$d$ hybridization. The calculated
trends agree with the measured data. For $N=5$ and 6 the positive
values are reported in \ct{Potzger02}, but with the remark that the
sign has not been measured but rather deduced from comparison to
calculations for $4sp$ impurities; thus here we show the same values
but also with different sign, according to our calculations. The Cd
$s$ moment is also shown, to demonstrate its proportionality to
$\hhf$. The trends are opposite for Te, where the on-site $s$-$p$
hybridization for lower $N$ splits the $s$-$d$ hybrids in two, one
populated and one unpopulated, for each spin. This means that the
parabolic decrease of $\hhf(N)$ measured for Cd is not necessarily
present for other impurities. Rather, each impurity has its own
characteristic trend, seen also in the data of Fig.~\ref{fig:1},
depending on the position of the hybrids.

We now discuss the limitations of the calculations. The local density
approximation has been known to underestimate the core electron
contribution to $\hhf$ in transition elements in the middle of the
$3d$ (Mn and Fe) and $4d$ series (Ru and Rh) \ct{Blugel87,Akai90}, but
otherwise it gives reasonable agreement with experiment for impurities
in bulk. Other terms than the Fermi contact interaction neglected in
the present work are the dipole and orbital moment terms. They can
contribute to $\hhf$ in the absence of cubic symmetry, and can be of
the order of a few percent compared to the contact term for $sp$
impurities at Fe/Ag interfaces \ct{Rodriguez01}; for Ni they should be
smaller due to the smaller magnetic moment. Finally, atom relaxations,
which are not included here, are known to affect $\hhf$ for impurities
in bulk Fe by a few percent, by changing the strength of the $s$-$d$
hybridization and shifting the antibonding states in the LDOS
\ct{Korhonen00}; they should have a similar effect in Ni.

In conclusion, using first-principles calculations we have interpreted
the measured~\ct{Potzger02} coordination dependence of $\hhf$ in Cd
impurities in Ni. Our results show a strong, nonlinear dependence on
the coordination number $N$, in agreement with the experiment.  We
have also predicted that the dependence of $\hhf(N)$ of the other
$5sp$ impurities in Ni is not decreasing as in Cd; rather, each
impurity exhibits its own ``fingerprint'' behavior. The reduced
coordination gradually lowers the antibonding $s$-$d$ hybrids in
energy, and more importantly reduces the symmetry causing an on-site
$s$-$p$ hybridization. Thus the trends of $\hhf$ for reduced
coordination are different than the ones in the bulk. The effects
should be similar for $sp$ impurities of the other lines of the
periodic table, and also for other ferromagnetic hosts. We hope that
our results will be stimulating for future experiments.

\begin{acknowledgments}
The author is grateful to Professor P. H. Dederichs and
Dr.~H. H. Bertschat for useful discussions. Financial support from the
Research and Training Network ``Computational Magnetoelectronics''
(contract RTN1-1999-00145) of the European Commission is gratefully
acknowledged.
\end{acknowledgments}

\end{document}